\documentclass[journal,10pt,twocolumn]{IEEEtran}
\usepackage{cite}
\usepackage[cmex10]{amsmath}
\usepackage{amsfonts}
\usepackage{algorithmic}

\usepackage{array}
\usepackage{mdwmath}
\usepackage{mdwtab}
\usepackage{bm}
\usepackage{multirow}

\usepackage{eqparbox}

\usepackage{graphicx}
\usepackage{booktabs}
\usepackage{algorithm}
\usepackage[tight,footnotesize]{subfigure}
\usepackage{caption}
\usepackage{cite}

\begin{document}
\bibliographystyle{ieeetr}
\captionsetup[figure]{name={Fig.},labelsep=period}

\title{Vehicle Tracking in Wireless Sensor Networks via Deep Reinforcement Learning}
\author{Jun Li, Zhichao Xing, Weibin Zhang, Yan Lin, and Feng Shu
\thanks{This work was supported in part by the National Natural Science Foundation of China under Grant 61727802 and Grant 61872184, in part by the Fundamental Research Funds for the Central Universities under Grant 30919011227, and in part by the Natural Science Foundation of Jiangsu Province under Grant BK20190454. \emph{(Corresponding authors: Yan Lin and Jun Li.)}}
\thanks{J. Li, Z. Xing, W. Zhang, Y. Lin, and F. Shu are with the School of Electronic and Optical Engineering, Nanjing University of Science and Technology, Nanjing 210094, China. (E-mail: \{jun.li, xing.zhichao, weibin.zhang, yanlin, shufeng \}@njust.edu.cn).}}
\maketitle

\begin{abstract}
Vehicle tracking has become one of the key applications of wireless sensor networks (WSNs) in the fields of rescue, surveillance, traffic monitoring, etc. However, the increased tracking accuracy requires more energy consumption. In this letter, a decentralized vehicle tracking strategy is conceived for improving both tracking accuracy and energy saving, which is based on adjusting the intersection area between the fixed sensing area and the dynamic activation area. Then, two deep reinforcement learning (DRL) aided solutions are proposed relying on the dynamic selection of the activation area radius. Finally, simulation results show the superiority of our DRL aided design.
\end{abstract}


\section{Introduction}
With the advances in the fabrication technologies that integrate the sensing and the wireless communication, a large-scale wireless sensor networks (WSNs) is formed in the desired fields by deploying dense tiny sensor nodes.
Vehicle tracking in WSNs has several prominent merits:
firstly, the sensing unit is close to the vehicle, thus the sensed data will be of a qualitatively good geometric fidelity via vehicle-to-infrastructure (V2I) technology \cite{4};
secondly, the information about the target is simultaneously generated by multiple sensors and thus contains redundancy \cite{5}.
However, there still exist some unsolved problems as far as vehicle tracking in WSNs concerned.
The first issue is how to guarantee the vehicle to be tracked by sensor nodes.
Second is how to maximize the tracking accuracy with the limited resources of WSNs, such as the energy restriction of each node \cite{9}.

There are two strategies based on the data processing mechanism, namely centralized strategy and decentralized strategy.
Specifically, in centralized strategy, a sensor is artificially selected as a cluster head, and the tracking estimation is performed at this node with all received data \cite{12}.
\begin{figure}[htbp]
\centering
\includegraphics[width=3.5in]{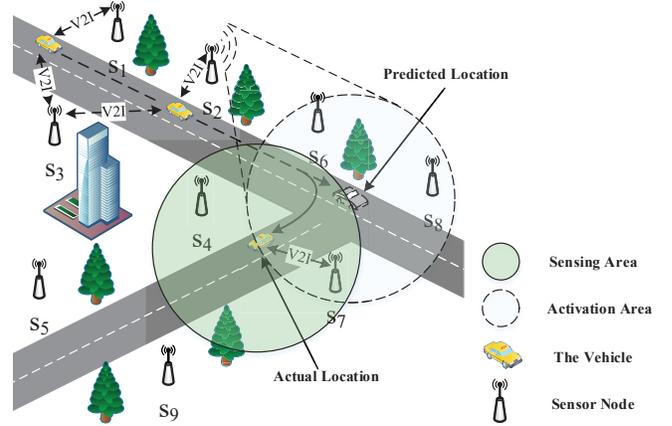}
\captionsetup{font={small}}
\caption{An illustration of an vehicle tracking scenario in WSNs.}
\label{fig. 1}
\end{figure}
However, the tasks at the cluster head may be overloaded in this strategy.
In each iteration of the decentralized strategy, each cluster uses the data from their neighbours to refine its local estimate \cite{main}.
Therefore, although the head is closer to the data source than the fixed head in centralized strategy, it has higher energy efficiency.
Meanwhile, the works in \cite{RL} point out that deep reinforcement learning (DRL) is a problem-solving tool and suitable for decentralized systems in WSNs.

The existing decentralized tracking strategies in WSNs are based on the prediction position to activate a fixed number of nodes, which will result in unnecessary energy consumption. Thus, we propose a dynamic activation area adjustment scheme based on DRL, to save energy consumption while ensuring tracking accuracy.
Based on this, we first formulate the problem as a Markov decision process (MDP).
Then, we construct the optimization problem as maximizing average rewards of MDP, where the reward consists of both tracking accuracy and energy consumption.
Furthermore, we propose a pair of schemes based on deep Q network (DQN) and on deep determined policy gradient (DDPG), to maximize average rewards of MDP.
In simulation, our proposed DQN and DDPG based algorithms outperform the conventional Q-learning based method in terms of tracking accuracy and energy consumption.

\section{System Model}
Fig. 1 illustrates an vehicle tracking scenario in a WSNs, which consists of a single vehicle and multiple sensor nodes.
The WSNs starts tracking and broadcasting when the object vehicle appears in its monitoring area.
Consider that the time slots are denoted by $t=1,2,...,T$, with unequally duration $\tau(t)$ seconds.
The vehicle derives into the monitoring area of the WSNs at time slot $t=1$, and leaves the WSNs or the WSNs loss vehicle at $t=T$.
Let the vehicle's vector denote as $\emph{\textbf{x}}(t) = [x_t,v_t^x,y_t,v_t^y]^{\text{T}}$, where $\emph{\textbf{l}}(t)=[x_t,y_t]^{\text{T}}$ is the 2-D position of the vehicle, and $[v_t^x,v_t^y]^{\text{T}}$ is the velocity of the vehicle at time slot $t$ on the plane.
As shown in Fig. 1, the connection between the sensor nodes and the vehicle can be established by V2I within a certain range, thus the radius $r_{\text{s}}$ of sensing area $n_{\text{s}}(t)$ is fixed.
Moreover, the WSNs activates the sensors in activation area $n_{\text{a}}(t)$ with radius $r_{\text{a}}(t)$ in advance, for tracking with dynamic energy consumption.
Additionally, let the sensors in set of intersection area $S_{\text{int}}(t) = n_{\text{s}}(t) \cap n_{\text{a}}(t)$ denote as $i = 1,2,\ldots,|m(t)|$, where $m(t)$ is the collection of sensors in $S_{\text{int}}(t)$, and $|m(t)| = \text{card}[m(t)]$ represents the amount of elements in set $m(t)$.

At time slot $t-1$, WSNs enables an activation area $n_{\text{a}}(t)$ in advance, which is an area based on the predicted vector $\hat{\emph{\textbf{x}}}'(t)$ and $r_{\text{a}}(t)$.
When the vehicle moves from $t-1$ to $t$, sensors in $m(t)$ track the vehicle cooperatively, and then only one sensor node is selected for transmitting data to the sensors in $n_{\text{a}}(t+1)$.

\subsection{Motion Model}
In our model, the vehicle's vector is 4-dimensional.
For simplicity, we model the vehicle's motion vector as a linear moving model \cite{model} as $
\emph{\textbf{x}}(t + 1) = \emph{\textbf{A}}\emph{\textbf{x}}(t) + \emph{\textbf{D}}\emph{\textbf{w}}(t)$,
where $\emph{\textbf{A}}$ represents the vector transition matrix, $\emph{\textbf{D}}$ is the transition duration matrix, and $\emph{\textbf{w}}(t)$ denotes the noise adding on the vehicle, which is a zero-mean white Gaussian noise with covariance matrix $\emph{\textbf{Q}}$.

\subsection{Measurement Model}
The vehicle's measurement model can be given by $
{\emph{\textbf{z}}_i}(t) = \emph{\textbf{H}}\emph{\textbf{x}}(t) + \emph{\textbf{v}}_i(t)$,
where ${\emph{\textbf{z}}_i}(t)$ represents the measurement of the $i$-th sensor at time $t$, and the measurement matrix $\emph{\textbf{H}}$ is assumed to be same for each sensor node.
$\emph{\textbf{v}}_i(t)$ is the measurement noise of the $i$-th sensor, which is assumed to be a zero-mean white Gaussian with covariance matrix $\emph{\textbf{R}}_i$.
Additionally, the observation noises that are added on different sensor nodes are independent.

\subsection{Filtering Model}
In this paper, we employ the Kalman filtering approach for WSNs to track the vehicle.
To be specific, the iterative process with motion and measurement model can be given by \cite{main} as
\begin{subequations}
\begin{equation}
\hat{\emph{\textbf{x}}}'(t) = \emph{\textbf{A}}\hat{\emph{\textbf{x}}}(t - 1) + \emph{\textbf{D}}\emph{\textbf{w}}(t-1),
\end{equation}
\begin{equation}
{\emph{\textbf{P}}'_i}(t) = \emph{\textbf{A}}\emph{\textbf{P}}_i(t - 1)\emph{\textbf{A}}^{\text{T}} + \emph{\textbf{Q}}(t - 1),
\end{equation}
\end{subequations}
where the priori estimated vector $\hat{\emph{\textbf{x}}}(t - 1)$ and its covariance $\emph{\textbf{P}}_i(t - 1)$ is obtained from the data in time slot $t-1$.
Then, the priori data is modified by $\hat{\emph{\textbf{x}}}'(t)$ and ${\emph{\textbf{z}}}(t)$:
\begin{subequations}
\begin{equation}
\emph{\textbf{P}}_i(t) = \left[ \emph{\textbf{P}}_i'(t)^{-1}
 + {\emph{\textbf{H}}}^{\text{T}}[\emph{\textbf{R}}_i(t)]^{-1}\emph{\textbf{H}}\right]^{-1},
\end{equation}
\begin{equation}
\hat{\emph{\textbf{x}}}(t) = \emph{\textbf{P}}_{\text{o}}(t) \left[\emph{\textbf{P}}_{\text{o}}'(t)^{-1} \hat{\emph{\textbf{x}}}'(t) + \emph{\textbf{H}}^{\text{T}} [\emph{\textbf{R}}_{\text{o}}(t)]^{-1} \emph{\textbf{z}}_{\text{o}}(t)\right].
\end{equation}
\end{subequations}
Herein, $\emph{\textbf{z}}_{\text{o}}(t)=\emph{\textbf{H}}\emph{\textbf{x}}(t) + \mathop {\rm{min}}\limits_{{i}} \emph{\textbf{v}}_i(t)$ denotes the optimal measurement by broadcasting the data of selected sensor nodes in $m(t)$, where $\emph{\textbf{R}}_{\text{o}}(t)=\mathbb{E}[\mathop {\rm{min}}\limits_{{i}} \emph{\textbf{v}}_i(t)]$.
The goal of this selection strategy can help improve tracking accuracy, namely the smaller $\emph{\textbf{P}}_i(t)$ is, the better estimation quality becomes\cite{trace}.
Next, we select a sensor with best measuring performance according to $\emph{\textbf{P}}_{\text{o}}(t)=\mathop {\rm{min}}\limits_{{i}} tr(\emph{\textbf{P}}_i(t))$,
where $\emph{\textbf{P}}_{\text{o}}(t)$ is the data that the fusion center needs to transmit to $n_{\text{a}}(t+1)$.

\subsection{Time Model}
In this paper, for obtaining the time interval of tracking vehicle in WSNs, we quantify the time based on the formula as \cite{uav}:
\begin{subequations}
\begin{equation}
\tau[c(\emph{\textbf{p}}_1, \emph{\textbf{p}}_2)] = \frac{N_b}{B\text{log}_2\left(1+\frac{P_{\text{t}} c(\emph{\textbf{p}}_1, \emph{\textbf{p}}_2)}{\sigma^2}\right)},
\end{equation}
\begin{equation}
c(\emph{\textbf{p}}_1, \emph{\textbf{p}}_2) = \left[\frac{\rho_0}{||\emph{\textbf{p}}_1-\emph{\textbf{p}}_2||_2}\right],
\end{equation}
\end{subequations}
where $\rho_0$ in (3\text{b}) denotes the path loss per meter, thus $c(\emph{\textbf{p}}_1, \emph{\textbf{p}}_2)$ is the channel gain between position $\emph{\textbf{p}}_1$ and $\emph{\textbf{p}}_2$.
In $(3\text{a})$, $N_{\text{b}}$ denotes the bits of each task, $B$ is the communication bandwidth, $P_{\text{t}}$ is transmission power, and $\sigma^2$ is power of Gaussian white noise in WSNs.
Therefore the data transmission duration is calculated as $\tau[c(\emph{\textbf{p}}_1, \emph{\textbf{p}}_2)]$.
Furthermore, the processing time consists of three parts:
\begin{enumerate}[(i)]
  \item Data Gathering Time Model: Based on the estimated vector $\hat{\emph{\textbf{x}}}(t-1)$, activation radius $r_{\text{a}}(t)$, and the sensing area $n_{\text{s}}(t)$, the sensors in $m(t)$ will detect the vehicle.
      Let $\emph{\textbf{p}}_1 = \hat{\emph{\textbf{l}}}(t)$, $\emph{\textbf{p}}_2 = \emph{\textbf{p}}(i), i=1,2,...,|m(t)|$, thus
      $\tau_1^i(t) = \tau[c(\hat{\emph{\textbf{l}}}(t), \emph{\textbf{p}}(i))]$, where the estimated position is $\hat{\emph{\textbf{l}}}(t) =\emph{\textbf{H}}\hat{\emph{\textbf{x}}}(t)$.
    In addition, let $\tau_1'(t) = \mathop {\rm{max}}\limits_{{i}}\tau[c(\hat{\emph{\textbf{l}}}(t), \emph{\textbf{p}}(i))]$ be the the longest duration, which based on the channel gain with the distance between the vehicle and the furthest node $i$.
  \item Data Fusion Time Model: After data gathering, the senors in $m(t)$ should establish a communication mechanism to find the optimal tracking data.
    Thus, the system randomly selects a node $J$ as a virtual data fusion center which is responsible for receiving data from other sensor node $j=1,2,...,|m(t)|$, $j \neq J$.
    Let $\emph{\textbf{p}}_1 = \emph{\textbf{p}}(J)$, $\emph{\textbf{p}}_2 = \emph{\textbf{p}}(j)$, thus $\tau_2^j(t) = \tau[c(\emph{\textbf{p}}(J), \emph{\textbf{p}}(j))]$.
     Let $\tau_2'(t) = \mathop {\rm{max}}\limits_{{j}}\tau[c(\emph{\textbf{p}}(J), \emph{\textbf{p}}(j))]$ represent the duration of total data fusion process, which based on the channel gain with distance between $J$ and furthest node $j$.
  \item Data Broadcast Time Model: After data fusion in $m(t)$, the center $J$ transfers the optimal data to the sensors in $n_{\text{a}}(t+1)$.
      Let $\emph{\textbf{p}}_1 = \emph{\textbf{p}}(J)$, $\emph{\textbf{p}}_2 = \emph{\textbf{p}}(k), k=1,2,...,|m_{\text{a}}(t+1)|$, thus $\tau_3^k(t) = \tau[c(\emph{\textbf{p}}(J), \emph{\textbf{p}}(k))]$.
    Similarly, $|m_{\text{a}}(t+1)|$ denote the amount of sensors of $n_{\text{a}}(t+1)$, and $\tau_3'(t) = \mathop {\rm{max}}\limits_{{k}}\tau[c(\emph{\textbf{p}}(J), \emph{\textbf{p}}(k))]$ denote the total data broadcast duration based on the minimum channel gain with the longest distance between $J$ and the sensor $k$.
\end{enumerate}

In summary, the time duration in each round includes these three parts $\tau(t)=\tau_1'(t)+\tau_2'(t)+\tau_3'(t)$.

\section{Problem Formulation}
This paper solves the problem of improving tracking accuracy and energy saving when tracking vehicle in WSNs.
Explicitly, the number of the sensor nodes for tracking contributes to the tracking accuracy \cite{main}.
Nevertheless, with the increase of the sensor nodes, the energy consumption for communication and data processing becomes higher.
To efficiently track the vehicle, we propose a dynamic activation area adjustment scheme to balance the trade-off between tracking accuracy and energy consumption.
In this section, we model this scheme as an MDP based on the nature of the dynamic decision-making problem.

\begin{itemize}
  \item State Space: Let $\mathcal{S}=\{s(t)|t=1,2...,T\}$ denote the state space of the WSNs, where $s(t) = \{\hat{\emph{\textbf{x}}}'(t),\hat{\emph{\textbf{x}}}(t),r_{\text{a}}(t)\}$ includes the estimated vector $\hat{\emph{\textbf{x}}}'(t)$ of the vehicle in time slot $t$, the estimated vector of vehicle $\hat{\emph{\textbf{x}}}(t)$ based on $\hat{\emph{\textbf{x}}}'(t)$ and the measurement ${\emph{\textbf{z}}_i}(t)$, and the radius of the activation area $r_{\text{a}}(t)$.
    The elements in $\mathcal{S}$ except $r_{\text{a}}(t)$ are described in the Kalman filtering model.
  \item Action Space: Let $\mathcal{A}$ denote the action space of the WSNs, where $r_{\text{a}}(t)\in\mathcal{A}$ is the radius of $n_{\text{a}}(t)$.
  \item Reward: Assuming that $n_{\text{a}}(t)$ and $n_{\text{s}}(t)$ are circles with radius of $r_{\text{a}}(t)$ and $r_{\text{s}}$, the prediction error in time slot $t$ is denoted as $ L = ||\hat{\emph{\textbf{l}}}'(t)-\hat{\emph{\textbf{l}}}(t)||_2 $, where the predicted position is $\hat{\emph{\textbf{l}}}'(t)= \emph{\textbf{H}}\hat{\emph{\textbf{x}}}'(t)$.
      In the process of the tracking, we consider the reward consisting of both the tracking accuracy and the energy consumption in each time slot:
  \begin{enumerate}[(i)]
  \item Tracking Accuracy: For quantifying the tracking accuracy to measure the tracking effect of WSNs, classical measurement (such as estimation error covariance) represents the error between $\hat{\emph{\textbf{l}}}'(t)$ and $\hat{\emph{\textbf{l}}}(t)$, but we would like to measure the error between $n_{\text{s}}(t)$ and $n_{\text{a}}(t)$, which depends on $\hat{\emph{\textbf{l}}}'(t)$ and $r_{\text{a}}(t)$.
  By adopting $S_{\text{int}}(t) = n_{\text{s}}(t) \cap n_{\text{a}}(t)$ as the metric, we can can measure the effect of $r_{\text{a}}(t)$ on energy consumption and track accuracy.
  As such, $\hat{\emph{\textbf{l}}}'(t) = \hat{\emph{\textbf{l}}}(t)$ satisfies when $L=0$ and $\text{max}[r_{\text{a}}(t)]=r_{\text{s}}$, thus as $S_{\text{int}}(t)$ increases, the predicted position becomes more accurate.
  We define the $
    S_{\text{int}}(t)  = r_{\text{a}}^2(t)[\theta_1(t)-\text{sin}\theta_1(t)\text{cos}\theta_1(t)] +  r_{\text{s}}^2[\theta_2(t)-\text{sin}\theta_2(t)\text{cos}\theta_2(t)] $.
    Herein, $\theta_1(t)$ and $\theta_2(t)$ are the angles of intersection area, which are obtained by $\theta_1(t) = \text{cos}^{-1}\frac{r_{\text{a}}^2(t)+L^2-r_{\text{s}}^2}{2r_{\text{a}}(t)L}$ and $\theta_2(t) = \text{cos}^{-1}\frac{r_{\text{s}}^2+L^2-r_{\text{a}}^2(t)}{2r_{\text{s}}L}$, respectively.

  \item Energy Consumption: In duration of $\tau(t)$, the energy consumption generated by data transmission is
    \begin{equation}
    \begin{split}
    e_{\text{com}}(t) = & P_{\text{r}} \sum\limits_{i=1}^{|m(t)|}\tau_1^i(t) + P_{\text{r}} \tau_2'(t) +  P_{\text{t}}\sum\limits_{j=1}^{|m(t)|-1}\tau_2^j(t) \\ & + P_{\text{t}} \tau_3'(t) + P_{\text{r}} \sum\limits_{k=1}^{|m_{\text{a}}(t+1)|}\tau_3^k(t),
    \end{split}
    \end{equation}
    where $P_{\text{t}}$ and $P_{\text{r}}$ denotes the power of nodes for transmitting and receiving data, respectively.
    In addition, the energy consumption of nodes for waking up in $n_{\text{a}}(t)$ and working in $m(t)$ is $
    e_\text{w}(t) = \left[P_{\text{w}} |m(t)| + P_{\text{idle}}(|m_{\text{a}}(t)| - |m(t)|)\right]\tau(t)$,
    where $|m_{\text{a}}(t)|$ denote the amount of sensors of $n_{\text{a}}(t)$, and the power of the nodes in working mode is denoted as $P_{\text{w}}$ and the power of the nodes in idle mode is denoted as $P_{\text{idle}}$\cite{main}.
    Thus, the total energy consumption is $e(t) = e_{\text{com}}(t) + e_{\text{w}}(t)$.
  \end{enumerate}
  Therefore, as the goal of this system is to improve the tracking accuracy and to minimize the energy consumption, we can define reward $R(t)$ by combining normalized $S_{\text{int}}(t)$ and $e(t)$ as $R(t) = \frac{S_{\text{int}}(t)}{\pi r^2_{\text{s}}} - \frac{e(t)}{e_{\text{max}}}$, where $e_{\text{max}}$ denotes the maximum energy consumption in each time duration.
\end{itemize}

It is noteworthy that, as the number of the training episodes increases, the total rewards in a single episode may exist a certain degree of disturbance, thus the increase of the total rewards in a single episode does not represent an accurate improvement of the system performance.
Therefore, our goal is to find an optimal policy $\pi^*$ with the maximum average total reward, given by:
\begin{equation}
\begin{split}
\pi^* = \arg\max \frac{\sum\limits_{t=1}^{T}{R(t)}}{T}.
\end{split}
\end{equation}

\section{Proposed DRL Based Methods}
In this section, to solve the problem (5), we adopt the policy iteration based DDPG method, and the value iteration based DQN method.
Additionally, in DQN method, we use the action selection strategy of 'softmax' to compare with the 'greedy' strategy.

\subsection{Proposed DDPG Based Method}
DDPG method is an algorithm which combines the Actor-Critic framework and the neural network\cite{ddpg}, where the Actor-Critic learns both the policy and value network.
In DDPG, when the state and action value are the input of a $\text{Q}$-function, it estimates a $\text{Q}$-value $Q^{\pi}(s,a)$ according to a policy $\pi$ as
$Q^{\pi^*}(s,a) \leftarrow \mathbb E_{s'}[r+\gamma\max\limits_{a'}Q^{\pi^*}(s',a')|s,a]$,
where $s$ and $a$ are the current state and action, $s'$ and $a'$ are the next state and action, and $\gamma\in[0,1]$ is a discount factor to the future reward.
The policy network is termed as the actor and the value network is termed as the critic.
In DDPG, the critic receives $(s, r)$ and produces a temporal-difference (TD) error $\sigma=r+\gamma Q(s',a')-Q(s,a)$.
With TD error, the actor network updates with the direction suggested by the critic network.
In DDPG based algorithm, for avoiding the influence of data correlation in training, the actor network is divided into a current actor network $\mu(s|\vartheta^\mu)$ and a target actor network $\mu'(s|\vartheta^{\mu'})$, while the critic network is divided into the current critic network $Q(s,a|\vartheta^Q)$ and the target critic network $Q'(s,a|\vartheta^{Q'})$.
The weights of four NNs will be updated as $\vartheta^{Q'}\leftarrow\ \iota\vartheta^Q + (1-\iota')\vartheta^{Q'}$ and $\vartheta^{\mu'}\leftarrow\ \iota\vartheta^\mu + (1-\iota')\vartheta^{\mu'}$, where $\iota$ is the learning decrement.

\subsection{Proposed DQN Based Method}

\begin{algorithm}[!t]
\caption{DQN Based Dynamic Activation Area Adjustment Scheme}
\begin{algorithmic}[1]
\STATE{Initialize $\vartheta$ and $\overline{\vartheta}$; $r$ and $q$; $\mathcal D$ and $N_t$; $\iota$ and $T=0$;}
\FORALL{episode $\ell=1,N_t$($N_t$ is the number of episodes)}
\STATE{Get the initial state $s_1$;}
\WHILE{the vehicle under tracking}
\STATE{Choose action based on $\varepsilon$-greedy policy or softmax policy;}
\STATE{Obtain reward $r$ and store $(s,a,r,s')$ in $\mathcal D$;}
\IF{the memory $\mathcal D$ is full}
\STATE{Sample mini-batch $K$ randomly from $\mathcal D$;}
\STATE{$\vartheta \leftarrow \min\limits_{\vartheta} l(\vartheta)$ and every $E$ steps reset $\widehat{Q}=Q$;}
\ENDIF
\STATE{Set $T = T + \tau(t)$, return q;}
\ENDWHILE
\ENDFOR
\label{code:recentEnd}
\end{algorithmic}
\end{algorithm}

This paper considers two action selection  methods:
\begin{itemize}
    \item $\varepsilon$-Greedy Policy: In each time slot, action selection according to $a' = \arg\max\limits_{a}Q(s,a;\vartheta)$ when        $\psi_t < \varepsilon$, and $\psi_t$ is a random variable. Then, select $a$ randomly in $\mathcal{A}$.
    At the end of each episode, the parameter $\varepsilon$ will be updated by $ \varepsilon = \text{max}\{\varepsilon_{\text{min}}, \varepsilon-\varsigma\}$,
        where $\varepsilon_{\text{min}}$ is the minimum for $\varepsilon$, and $\varsigma$ is the decrease speed.

    \item Softmax Policy: For getting the selection probability $P(a)$ of action $a$, the softmax based action selection policy is formulated as $P(a)=\frac{\text{e}^{\frac{1}{t}\sum_{1}^{t}r(a)}}{\sum_{i=1}^{\text{card}(\mathcal{A})}{\text{e}^{\frac{1}{t}\sum_{1}^{t}r(a_i)}}}$, where $r(a)$ is the reward based on action $a$, $r(a_i)$ is the reward based on  current action $a_i$.
\end{itemize}

Additionally, DQN employs the two networks with the same structure but different parameters $\vartheta$ and $\bar{\vartheta}$ respectively.
For training the model, we search the weights by minimizing the loss function $
\min\limits_{\vartheta} l(\vartheta) = \sum \limits_{t=0}^T[\hat{Q}(s,a;\vartheta)-y_t^{\text{DQN}}]^2$, where $
y_t^{\text{DQN}} = r_t + \gamma\max\limits_{a'}\hat{Q}(s',a';\bar{\vartheta})$,
$\hat{Q}(s,a;\vartheta)$ represents a predicted maximum Q-value with weight $\vartheta$, and $y_t^{\text{DQN}}$ represents the target Q-value  with weight $\bar{\vartheta}$.
The details of this process are shown in Algorithm 1.

\section{Simulations and Results}

\begin{table}[h]
\small
\caption{Simulation Parameters}
\begin{tabular}{cccccc}
\toprule  
parameter & value & parameter & value & parameter & value\\
\midrule  
$B(\text{MHz})$                 & $1$             & $N_{\text{b}} (\text{bits}) $         & $2 \times 10^{4}$  & $P_{\text{w}}  (\text{w})$ & $5$ \\
$P_{\text{idle}}(\text{w})$ & $0.05$  & $\gamma$      & $0.9$               & $\mathcal D$   & $2000$\\
$N_t$                    & $1800$            & $\vartheta$   & $0.5$               & $E$            & $500$\\
$K$                      & $30$              & $\varepsilon$ & $0.9$              & $\iota'$       & $0.01$\\
$\vartheta^{\mu}$                      & $0.5$              & $\vartheta^{Q}$ & $0.5$              & $\vartheta^{\mu'}$         & $0.5$\\
$\vartheta^{Q'}$ & $0.5$ & $\bar{\vartheta}$ & $0.5$   &$\sigma^2$(dBm)  & -110 \\
\bottomrule 
\end{tabular}
\end{table}

Based on the previous research in \cite{model}, we consider the field of WSNs is $360\times600\text{m}^2$.
Besides, the motion model is
\begin{figure}[htb]
\centering
\includegraphics[width=3.5in,angle=0]{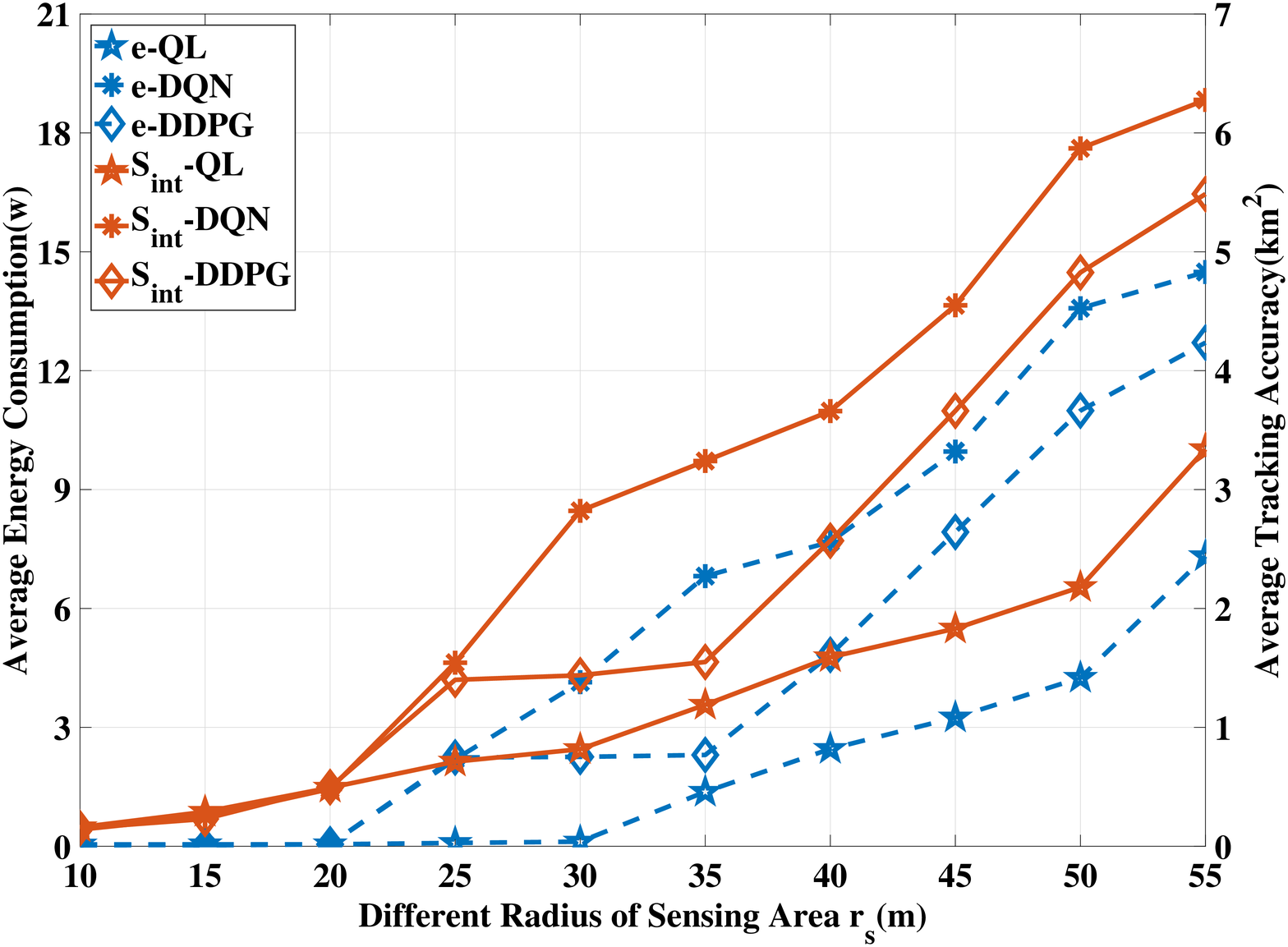}
\captionsetup{font={small}}
\caption{Energy consumption and tracking accuracy versus $\text{r}_{\text{s}}$.}\label{fig:Different schemes}
\end{figure}
\begin{equation}
\emph{\textbf{x}}(t+1) = \begin{bmatrix}1 & \tau & 0 & 0\\ 0 & 1 & 0 & 0\\ 0 & 0 & 1 & \tau\\ 0 & 0 & 0 & 1\end{bmatrix}\emph{\textbf{x}}(t) + \begin{bmatrix} \frac{\tau^2}{2} & 0\\ \tau & 0\\ 0 & \frac{\tau^2}{2}\\ 0 & \tau \end{bmatrix} \begin{bmatrix} w_x(t) \\ w_y(t)\end{bmatrix},
\end{equation}
where $\emph{\textbf{x}}(t)$ is the vector of the vehicle, $\tau = \tau(t)$ is the length of each time interval, and $w(t)$ is the vector noise.
The initial vector is $\emph{\textbf{x}}(0) = [0\ 7.84\ 0\ 7.84]^{\text{T}}$, which means the initial speed is 40 kmph.
The vector noise accounting for the unpredictable modeling error is characterized by $
\emph{\textbf{Q}} = \mathbb{E}[\emph{\textbf{w}}(t)\emph{\textbf{w}}^{\text{T}}(t)] = \begin{bmatrix}0.03 & 0\\ 0 & 0.03\end{bmatrix} $.

In the tracking of a single vehicle, there are many sensors that can be used in the scenario, such as cameras \cite{main} and logical combination of ranging sensors \cite{5}.
By adopting the strategy of uniform distribution of sensor nodes in \cite{main}, we set 375 nodes in the sensing range with the density of 1 sensor/24$\text{m}$, and the linear measurement model of each sensor is $
\emph{\textbf{z}}_i(t) = \emph{\textbf{H}} \emph{\textbf{x}}(t) + \begin{pmatrix} v_x(t) \\ v_y(t)\end{pmatrix}_i $,
where $\emph{\textbf{H}} = \begin{bmatrix}1 & 0 & 0 & 0\\ 0 & 0 & 1 & 0\end{bmatrix}$.
Notably, the motion model can also be a non-linear model, which can be estimated by extended Kalman filter, particle filter or other methods.
The parameters of the measurement noise covariance are $
\emph{\textbf{R}}_i = \mathbb{E}\left[ \begin{pmatrix} v_x(t) \\ v_y(t)\end{pmatrix}_i\ \begin{pmatrix} v_x(t) \\ v_y(t)\end{pmatrix}_i^{\text{T}}\right] = \begin{bmatrix}400 & 0\\ 0 & 400\end{bmatrix} $.
We randomly initialize $\emph{\textbf{x}}(0)$, and the parameters used in this model are shown in Table 1.

Fig. 2 plots the trends of the average tracking accuracy $\frac{1}{T}\sum\limits_{t=1}^{T}{S_{\text{int}}(t)}$ and the average energy consumption $\frac{1}{T}\sum\limits_{t=1}^{T}{e(t)}$ with different $r_{\text{s}}$, where each value of $e(t)$ and $S_{\text{int}}(t)$ is the final convergence value obtained by training 1800 episodes.
Clearly, increasing sensing area $n_{\text{s}}(t)$ achieves higher $S_{\text{int}}(t)$ and $e(t)$.

Fig. 3 shows the average accumulated rewards $\frac{1}{T}\sum\limits_{t=1}^{T}{R(t)}$ of the proposed algorithms.
Firstly, we can observe that although $e(t)$ and $S_{\text{int}}(t)$ increase with $r_{\text{s}}$, their gaps in terms of the average accumulated rewards enlarge gradually.
Secondly, as episodes go by, the average accumulated rewards based on DQN is much higher than other methods. The reasons include two aspects: first the QL based method leads to the lowest value because of the Q-table has a limited capacity; second the value iteration of DQN is based on the unbiased estimation of data, while the DDPG is based on the biased estimation.

\begin{figure}[htb]
\centering
\includegraphics[width=3.5in,angle=0]{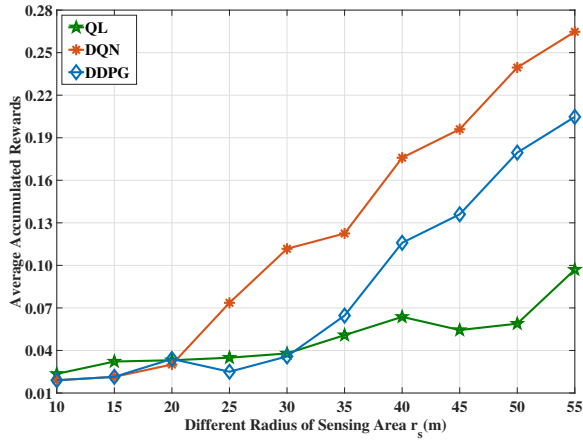}
\captionsetup{font={small}}                  
\caption{Average accumulated rewards versus $\text{r}_{\text{s}}$.}\label{fig:Different schemes}
\end{figure}

\section{Conclusions}
This paper focuses on improving the tracking accuracy and energy saving of target vehicle in WSNs.
Thus, we propose a decentralized vehicle tracking strategy, which dynamically adjusts the activation area to improve tracking accuracy and energy saving.
Then, we define the problem as a MDP and use DRL method to solve it.
Finally, the simulation results show that the DQN-based method has a better performance than other DRL-based methods.
In future, we can explore multi-agent DRL algorithms in cooperative tracking scenarios.

\bibliographystyle{IEEEtran}
\bibliography{myreference}

\end{document}